\documentclass[sigconf]{acmart}
\usepackage{bm}
\usepackage{amsmath}
\usepackage{multirow}
\usepackage{natbib}

\AtBeginDocument{%
  \providecommand\BibTeX{{%
    \normalfont B\kern-0.5em{\scshape i\kern-0.25em b}\kern-0.8em\TeX}}}
    
\copyrightyear{2021}
\acmYear{2021}
\setcopyright{rightsretained}
\acmConference[WSDM '21]{Proceedings of the Fourteenth ACM International Conference on Web Search and Data Mining}{March 8--12, 2021}{Virtual Event, Israel}
\acmBooktitle{Proceedings of the Fourteenth ACM International Conference on Web Search and Data Mining (WSDM '21), March 8--12, 2021, Virtual Event, Israel}\acmDOI{10.1145/3437963.3441769}
\acmISBN{978-1-4503-8297-7/21/03}

\begin{document}
\fancyhead{}

\title{Leave No User Behind: Towards Improving the Utility of Recommender Systems for Non-mainstream Users}

%
\author{Roger Zhe Li}
\affiliation{%
  \institution{Delft University of Technology}
  \city{Delft}
  \country{The Netherlands}
}
\email{Z.Li-9@tudelft.nl}

\author{Juli\'an Urbano}
\affiliation{%
  \institution{Delft University of Technology}
  \city{Delft}
  \country{The Netherlands}
}
\email{J.Urbano@tudelft.nl}

\author{Alan Hanjalic}
\affiliation{%
  \institution{Delft University of Technology}
  \city{Delft}
  \country{The Netherlands}
}
\email{A.Hanjalic@tudelft.nl}

%
\renewcommand{\shorttitle}{Leave No User Behind}
%
\sloppy
\begin{abstract}
In a collaborative-filtering recommendation scenario, biases in the data will likely propagate in the learned recommendations. In this paper we focus on the so-called mainstream bias: the tendency of a recommender system to provide better recommendations to users who have a mainstream taste, as opposed to non-mainstream users. We propose NAECF, a conceptually simple but effective idea to address this bias. The idea consists of adding an autoencoder (AE) layer when learning user and item representations with text-based Convolutional Neural Networks. The AEs, one for the users and one for the items, serve as adversaries to the process of minimizing the rating prediction error when learning how to recommend. They enforce that the specific unique properties of all users and items are sufficiently well incorporated and preserved in the learned representations. These representations, extracted as the bottlenecks of the corresponding AEs, are expected to be less biased towards mainstream users, and to provide more balanced recommendation utility across all users. Our experimental results confirm these expectations, significantly improving the recommendations for non-mainstream users while maintaining the recommendation quality for mainstream users. Our results emphasize the importance of deploying extensive content-based features, such as online reviews, in order to better represent users and items to maximize the de-biasing effect. 
\end{abstract}

%
%
%


\maketitle

\section{Introduction}

Collaborative Filtering (CF) models are the most investigated and deployed models in the domain of recommender systems~\citep{adomavicius2005toward}. These models assume that users who had similar preferences on items of a specific kind (e.g. books, movies) in the past may continue having similar preferences on other items of the same kind. The preferences of the users are expressed
through their explicit (e.g. ratings) or implicit (e.g. clicks, downloads) interactions with items. 

Among the CF models, \emph{Matrix Factorization} (MF)~\citep{koren2009matrix}, which tries to find a representation of both users and items in the same latent factor space, has long been the most successful and most widely deployed CF model. More recently, generalized factorization models, such as factorization machines~\citep{rendle2010factorization}, have been proposed, exploiting input beyond user-item interactions to learn the latent space. Exploiting more input, such as contextual features and other types of useful side information about users and items, was shown to further improve the recommendation quality. The potential for further improvement, for instance by relying on more abundant input data including audio, visual and textual item descriptions or social network dynamics, came only when deep neural networks (DNN)~\citep{goodfellow2016deep} entered the recommendation domain and enabled more sophisticated user/item representation space learning. In particular, textual data acquired from websites have been extensively exploited for this purpose, allowing users to leave review comments for items along with ratings. For this type of data, DNN-based user/item modeling utilizing NLP techniques has been shown to achieve significantly higher recommendation performance~\citep{zheng2017joint, cheng20183ncf, DBLP:conf/kdd/TayLH18, yang2020learning} as well as provide convincing explanations \citep{chen2018neural, cheng2019mmalfm, wang2018explainable, DBLP:conf/ijcai/PanLLZ20}. 

While these developments have greatly contributed to the improvement of the overall recommendation accuracy, one problem has remained largely unsolved, namely the presence of various types of biases in the learned recommendation models. In this paper we focus on the bias towards the so-called \emph{mainstream users}. A mainstream user often prefers items liked by many people and also reacts negatively to items widely disliked by others~\citep{schedl2018analysis}. Contrary to this, non-mainstream users typically show interest on rarely-visited items or have an opposite attitude towards widely accepted or rejected items. Such a ``grey sheep'' property~\citep{DBLP:conf/adma/ZhengAS17} makes these users different from others, making it difficult for a CF algorithm to identify similar peers. This leads to recommendations of a generally lower quality, because recommendations for these users are based on neighbors with insufficiently similar preferences. Furthermore, non-mainstream users are typically a minority and the numerous consistent user-item interactions within the cluster of mainstream users are likely to be dominant in steering the process of learning the user/item representation space. Because of this, the non-mainstream users and their preferred (``outlier'') items become underrepresented in such a space, leading to inequality of the recommendation performance across the user population. This is the \emph{mainstream bias}, the tendency to provide better recommendations to the mainstream users. 
Such bias could make non-mainstream users draw insufficient utility from a recommender system and could discourage them from using it anymore. This could lead to online businesses starting to lose customers. For the information and news recommender systems, however, we foresee even more serious consequences. Recommender systems may namely become less inclusive with respect to non-mainstream opinions and (e.g. political) views and in this way contribute to undesired long-term effects, like intellectual segregation and societal polarization.

In this paper we build on the success of DNNs for developing recommendation models and propose a simple but effective solution towards neutralizing the mainstream bias. With our new recommendation model, referred to as \textbf{N}eural \textbf{A}uto\textbf{E}ncoder \textbf{C}ollaborative \textbf{F}iltering (NAECF), we introduce adversarial conditions to the process of learning the recommendation algorithm, which in this specific case is realized as the minimization of the rating prediction error. The adversarial conditions are imposed by 
\emph{autoencoders}~\citep{ballard1987modular}, a deep learning architecture widely used for recommendation~\citep{sedhain2015autorec, okura2017embedding, ma2019gated}, added to a state-of-the-art DNN-based recommendation framework. They enforce that the user and item representations are learned in a way such that they preserve their specific and unique properties before being fed to the rating predictor. 
Since this preservation is achieved for all users, mainstream or not, the autoencoders prevent that the learned representations are biased towards the users with a mainstream taste. The results of experiments conducted on different domains and scales of the real-world datasets from Amazon~\citep{mcauley2015image} 
show that the representations learned in this way indeed help to de-bias the produced recommendations (predicted ratings in this case). Compared to the case without deploying the adversarial conditions, our proposed method produces significantly better recommendations for non-mainstream users while largely maintaining  the recommendation quality for mainstream users. We clearly show that this performance improvement largely stems from adding adversarial conditions to the process of user and item representation learning. In addition, our experiments demonstrate the benefit the non-mainstream users draw from the application of content-based features such as online reviews, further highlighting their value for achieving high recommendation quality across the user community.

The proposed NAECF approach is, to the best of our knowledge, the first to enforce preserving the unique user and item properties as the adversary to the process of learning how to recommend. This allows us to effectively address the mainstream bias in recommender systems, which has not been extensively studied this far.


A recommender system can be designed to either predict ratings or rank items. Despite the latter is picking up momentum in the field, we still choose in this paper to follow the rating prediction paradigm. The main reason for this lies in the core of our contribution, which is to investigate how a state-of-the-art recommendation framework may be extended in order to de-bias the process of generating recommendations. Since the framework we build upon was evaluated in terms of rating prediction, we follow this same paradigm in this paper.
Nonetheless, the user-item representation space generated by the autoencoders can serve to predict both ranking and ratings, so we do not consider our choice to limit the broad application of our proposal to the recommendation practice. 

The results of the paper can be fully reproduced with data and code available online\footnote{\url{https://github.com/roger-zhe-li/wsdm21-mainstream}}.

\section{Related work}

Our work relates mainly to two topics: biases in recommender systems and review-based user/item modeling.


\subsection{Biases in Recommender Systems}

Potential biases in the training data have already been recognized in early work on matrix factorization for recommendation. \citet{koren2009matrix} introduced a correction in the dot-product rating prediction formula to incorporate rating biases across users, that is, how the rating scale is interpreted by different users. Another bias related to ratings is the anchoring bias; it emerges from the influence of previous recommendations to a user on that user's future ratings. \citet{adomavicius2014biasing} explored two approaches to neutralize this bias. The first approach involves computational post-hoc adjustments of the ratings that are known to be biased. The second approach involves a user interface by which the system tries to prevent this bias \emph{during} rating. A different sort of bias is the popularity bias, due to which popular items may be recommended more frequently than other, less popular items (e.g., long-tail). \citet{abdollahpouri2018popularity} proposed an add-on to a general collaborative filtering algorithm by which a trade-off between accuracy and long-tail coverage can be tuned.
More recently, the discussion about biases has increasingly been conducted in the context of resolving ethical and societal issues when deploying recommender systems in practice, such as polarization~\citep{DBLP:conf/wsdm/RastegarpanahGC19}, fairness~\citep{DBLP:conf/cikm/MehrotraMBL018} and discrimination~\citep{DBLP:conf/flairs/MansouryASDPM20}%
, giving a further boost to the research on this topic. 

In this paper we focus on the aforementioned mainstream bias. While being conceptually close to popularity bias~\citep{DBLP:conf/recsys/AbdollahpouriMB19,DBLP:conf/ecir/KowaldSL20}, there is an important difference between the two. Popularity bias could lead to a separation between more and less popular items, similar to the separation of items into those being interesting to mainstream and non-mainstream users. However, popularity bias is not informative regarding the way a recommender system serves different groups of users. According to \citet{steck2011item}, users may tend to provide feedback on popular items simply by following (being influenced by) other users. In this way, their preferences are likely to be unconsciously driven away from their real interest. By focusing on mainstreamness, we explicitly look at the bias in the user population. 

\citet{DBLP:conf/ecir/KowaldSL20} demonstrated that non-mainstream music listeners are likely to receive the worst recommendations.  \citet{DBLP:journals/corr/abs-1912-11564} investigated music preferences across age groups. They observed that, although only taking a small proportion of users, kids and adolescents have significantly different preferences from other age groups in terms of music genres, and the recommendation performance on these two groups is also distinctive among all users. To repair the unfairness caused by the mainstream bias, several recent works aim at identifying non-mainstream music listeners and using the power of cultural aspects~\citep{schedl2018analysis, mullner2019studying} and human memories~\citep{DBLP:journals/corr/abs-2003-10699} to better profile these underrepresented users in recommender systems. Despite the reported progress, existing methods to alleviate the mainstream bias usually rely on their specific definitions of mainstreamness, which may limit the findings. Furthermore, these methods tend to split users into different mainstreamness groups for individual training. This setting may cause the loss of recommendation accuracy due to not exploiting the cross-group collaborative information. The approach proposed in this paper aims at neutralizing the mainstream bias in a more generic fashion and without divisions within the user population.



\subsection{Review-based User/Item Modeling} 
Supported by the rapid development of natural language processing (NLP) techniques, online reviews have increasingly been identified as an important source of useful information for addressing data sparsity issues in recommendation. Exploiting these reviews has led to several advanced recommendation concepts, pioneered by Collaborative Deep Learning (CDL)~\citep{wang2015collaborative}. This concept introduces a hierarchical Bayesian model using Stacked Denoising Autoencoders (SDAE) to reconstruct the rating matrix from encoded textual reviews. Another method, DeepCoNN~\citep{zheng2017joint}, unifies the processes of learning the user/item representation and rating prediction in an end-to-end model. 
The unification is achieved through a combination of Convolutional Neural Networks (CNN) and factorization machines.
Due to the sequential nature of reviews, Recurrent Neural Networks (RNN) and attention models are also widely used for user and item feature learning. \citet{wu2016joint} trained the review representations and ratings jointly within a Long Short-Term Memory (LSTM) framework for movie recommendation. \citet{chen2018neural} 
extended the DeepCoNN concept by incorporating attention factors into NARRE, a DeepCoNN-based framework, to provide convincing explanations. MPCN~\citep{DBLP:conf/kdd/TayLH18} is another attention-based model, which uses two hierarchical attention layers to infer the review importance. Although the use of text reviews has partially resolved the data sparsity issues, a more direct way to achieve this is to increase the scale of the training data. As an example, AugCF~\citep{DBLP:conf/kdd/WangYWNHC19} was proposed on top of DeepCoNN to augment review and rating data using Generative Adversarial Networks~\citep{DBLP:conf/nips/GoodfellowPMXWOCB14}. All models mentioned above represent users and items following the same principle, and the representations are derived from the same data source. Contrary to this, NPA~\citep{DBLP:conf/kdd/WuWAHHX19} and NeuHash-CF~\citep{DBLP:conf/sigir/Hansen0SAL20} represent users and items in different ways. While they model the items using content-based information, the users are represented by one-hot coded userID.
 
Despite the remarks expressed in literature state that reviews serve the recommendation better as regularizers than features~\citep{DBLP:conf/sigir/SachdevaM20}, the models mentioned above have been reported to achieve remarkable overall recommendation accuracy, showing the benefit of using textual review data as input.
In this paper, we look at online reviews from a different angle and further than accuracy alone. We analyze their value in achieving better user representations that allow us to balance the recommendation quality across users. We show that, with our proposed recommendation model, reviews can be instrumental in neutralizing the mainstream bias.

\section{Proposed Model: NAECF}

\begin{figure}[t]
\includegraphics[width=\columnwidth]{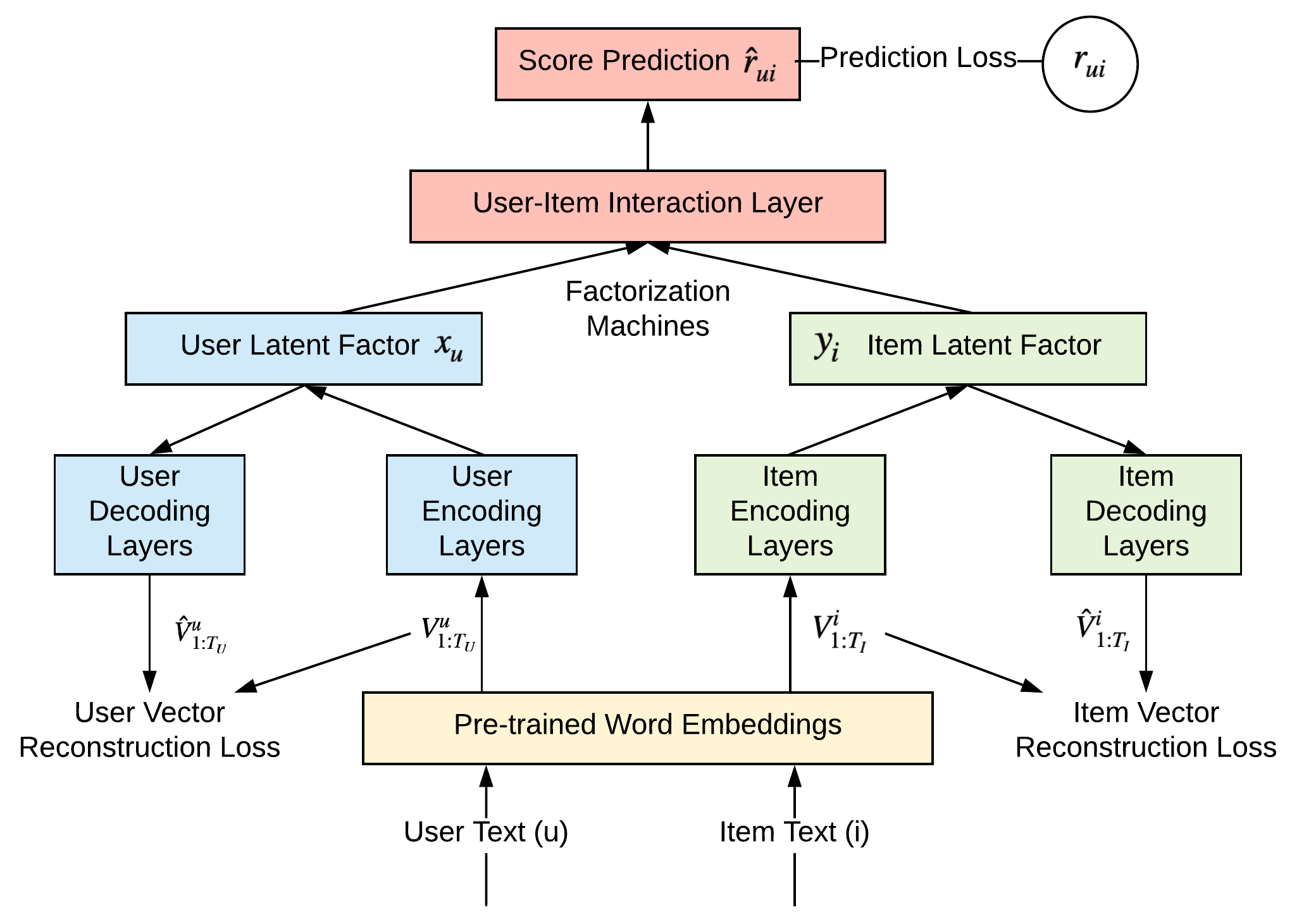}%
\caption{Overall architecture of NAECF.}%
\label{architecture}
\end{figure}

The architecture of the proposed NAECF model is illustrated in Fig.~\ref{architecture}. The scheme shows that with NAECF we pursue two learning goals simultaneously: maximizing the recommendation accuracy and reconstructing the users' and items' original feature vectors in the autoencoders. These feature vectors consist of the texts of user reviews, so we refer to the process taking place in the two AEs as ``text reconstruction''. Recommendation accuracy may be achieved by optimizing for rating prediction or ranking prediction. 
Since DeepCoNN~\citep{zheng2017joint}, the strongest baseline for comparison, is designed for rating prediction, we also take rating prediction as the criterion for recommendation optimization. This allows us to assess specifically the effect of enforcing user and item reconstruction as an adversarial condition to recommendation optimization on the mainstream bias. If the effect is there, it can also be expected if a ranking prediction scheme is expanded in the same way.

\subsection{Model Formulation}

The data we use consist of tuples $(u, i, r_{ui}, c_{ui})$, representing a user $u$ providing a rating $r_{ui}$ to item $i$ and leaving a review text $c_{ui}$ for said item.
Based on Fig.~\ref{architecture}, we see the realization of the overall goal of NAECF by minimizing the following loss: 
\begin{equation}
L = L_R + w\left(L_U + L_I\right)~, \label{eq:overall_loss}
\end{equation}
where $L_R$, $L_U$ and $L_I$ are, respectively, the mean rating prediction loss, and the mean text reconstruction losses for users and items. The constant $w$ is a weight determining the relative influence of user and item AEs compared to the rating prediction module. The three losses are defined by the following expressions:
\begin{align}
L_R &=\frac{1}{N_R}\sum_{u,i}{loss_R(u,i)} \label{loss}\\
L_U &=\frac{1}{N_U}\sum_{u}{loss_U(u)} \\
L_I &=\frac{1}{N_I}\sum_{i}{loss_I(i)}~,
\end{align}
where $N_R$, $N_U$ and $N_I$ represent the number of interactions, users and items in the training set, respectively. Normalizing by these terms makes the effect of the weight $w$ invariant to the statistics of the dataset.

\subsection{Learning for Rating Prediction}
In NAECF, the rating prediction loss for an individual user-item interaction is computed as a traditional squared loss
\begin{equation}
    loss_R (u,i) = \left(\frac{r_{ui} - \hat{r}_{ui}}{r_{max}-r_{min}}\right)^2,
\end{equation}
where $\hat{r}_{ui}$ is the predicted rating given by user $u$ to item $i$. The loss is normalized by the limits of the rating scale used in the dataset, so that $L_R$ is bounded between 0 and 1. 
The prediction is computed for the interaction $\hat{z}=(\bm{x}_u,\bm{y}_i)$ between vectors $\bm{x}_u$ and $\bm{y}_i$, encoding the user and item text feature representations, respectively. $\bm{x}_u$ and $\bm{y}_i$ are the latent factors, low-rank representations of users and items, extracted as the bottlenecks of the corresponding AEs, as indicated by the green and blue blocks in Fig.~\ref{architecture}. We follow the settings of DeepCoNN~\citep{zheng2017joint} with a Factorization Machine layer~\citep{DBLP:journals/tist/Rendle12}, and compute rating prediction as
\begin{equation}
		\hat{r}_{ui}=
			\hat{a}_0 +
			\sum\limits_{m=1}^{|\hat{z}|}\hat{a}_m\hat{z}_m +
			\sum\limits_{m=1}^{|\hat{z}|}\sum\limits_{n=m+1}^{|\hat{z}|}
				\langle\bm{\hat{v}}_m, \bm{\hat{v}}_n\rangle\hat{z}_m\hat{z}_n
\end{equation}
where $\hat{a}_0$ denotes the global bias and $\hat{a}_m$ denotes the strength of first order interactions in $\hat{z}$. Second order interactions are modeled by $\langle\bm{\hat{v}}_m, \bm{\hat{v}}_n\rangle = \sum_{f=1}^{|\hat{z}|}{\mathbf{\hat{v}}_{m,f}\mathbf{\hat{v}}_{n,f}}$ .

\begin{figure}[t]
\includegraphics[width=\columnwidth]{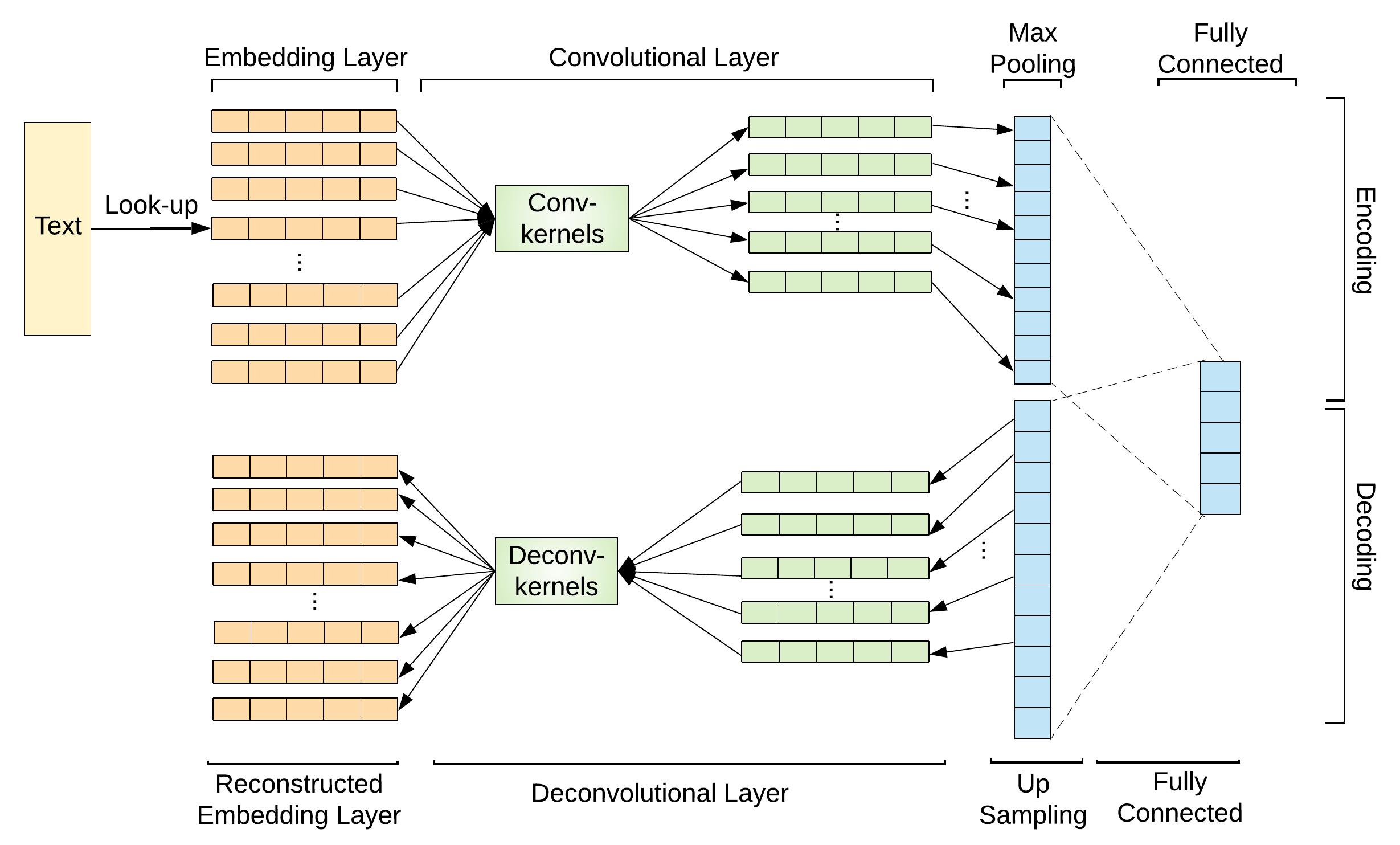}%
\caption{Architecture of the convolutional autoencoder for text feature transformation and extraction.}%
\label{text_autoencoder}
\end{figure}

\subsection{Learning for Textual Feature Transformation and Reconstruction}
The latent factors $\bm{x}_u$ and $\bm{y}_i$ are used not only for rating prediction, as indicated in the previous section, but also to reconstruct the original user and item representations in the computation of text reconstruction losses. We use an encoder to generate latent factors, which takes an initial user representation $V_u$ or item representation $V_i$. We deploy the strategy proposed by DeepCoNN~\citep{zheng2017joint}, that applies TextCNNs~\citep{DBLP:conf/emnlp/Kim14} for feature transformation.
For an arbitrary user $u$, we extract all review texts they authored and concatenate them into a single long document. Similar to the top NLP models like BERT~\citep{DBLP:conf/naacl/DevlinCLT19} and GPT-2~\citep{radford2019language}, here we adopt a cutoff length $T_U$ to truncate words exceeding the limit. For users with fewer than $T_U$ words, we pad empty words, denoted by \texttt{<UNK>} and initialized by zeros. This way all users are represented by the same number of review tokens. Then we introduce a look-up layer to get the initial individual word embeddings from a pre-trained model. By concatenating them, we obtain user embedding ${V_{1:T_U}^u}$. Similarly, we obtain item embedding ${V_{1:T_I}^i}$ for item $i$.


Encoding these initial ${V_{1:T_U}^u}$ and ${V_{1:T_I}^i}$ embeddings results in the latent factors, which then serve as input to a decoder that we introduce to create reconstructed embeddings $\hat{V}^u_{1:T_U}$ and $\hat{V}^i_{1:T_I}$. The architecture of the decoder is symmetric to the encoder with deconvolution and unpooling layers, as shown in Fig.~\ref{text_autoencoder}.
All hyper-parameters used in the decoding stage are the same as in the encoding stage.

The success of reconstructing initial user and item embeddings is modeled by the text reconstruction losses, which are computed for each user and item. 
In order to have scores on a bounded scale,
we rely on the cosine similarity to measure text reconstruction loss. Unlike most cases in text analysis where embedding values are positive, the original pre-trained embeddings we use in this paper do have negative values, making the cosine similarity range from -1 to 1. Therefore, we also normalize cosine similarities so that the scales of $L_U$ and $L_I$ are comparable to that of $L_R$. This leads to the following formulation of the individual text reconstruction losses: 
\begin{align}
\label{eq:tr_loss}
loss_U(u) &= \Bigg(\frac{1 - cos\big(V_{1:T_U}^u, \hat{V}_{1:T_U}^u\big)}{2}\Bigg)^2~, \\
loss_I(i) &= \Bigg(\frac{1 - cos\big(V_{1:T_I}^i, \hat{V}_{1:T_I}^i\big)}{2}\Bigg)^2~,
\end{align}
where $cos$ stands for the cosine similarity between the original vectors and the reconstructed ones.

\subsection{Model Learning}

We use Adaptive Moment Estimation (Adam)~\citep{kingma2014adam} to minimize the overall loss function in Eq.~(\ref{eq:overall_loss}). This way, training converges fast and the learning rate is adapted during the process.
\section{Experimental Design}

Here we present a series of experiments designed to evaluate the proposed NAECF model through the following research questions:
\begin{itemize}
\item RQ1: Does NAECF improve the recommendation for non-mainstream users, creating a better balance across users?
\item RQ2: What is the effect of using reviews and textual feature transformations on mainstream and non-mainstream users?
\item RQ3: What is the correlation between recommendation accuracy and the difficulty of user feature reconstruction?
\end{itemize}


\subsection{Data and Metrics}

\begin{table}[t]
  \caption{Statistics of the datasets.}\label{stat}%
  \centering{\footnotesize\begin{tabular}{|rrrrrr|}\hline
  Dataset & \#users &\#items & \#ratings & Sparsity & \#words \\ \hline
  Instant videos & 5,130 & 1,685 & 37,126 & 99.57 & 19M \\ 
  Digital music & 5,541 & 3,568 & 64,706 & 99.67 & 73M \\
  BeerAdvocate & 3,703 & 37,580 & 393,035 & 99.72 & 198M \\ \hline
  \end{tabular}}
\end{table}

In this paper we focus on improving the recommendation for non-mainstream users, and investigate the power of text reviews for this purpose. Therefore, the selected datasets are all review-based (see Table~\ref{stat}). We use two Amazon real-world datasets\footnote{\url{http://jmcauley.ucsd.edu/data/amazon/}} covering different recommendation domains, namely instant videos and digital music, and another dataset from BeerAdvocate~\citep{mcauley2012learning}\footnote{\url{http://snap.stanford.edu/data/web-BeerAdvocate.html}}. The ratings all range from 1 to 5. However, in the Amazon datasets ratings are integers, while in the BeerAdvocate dataset they are multiples of 0.5. Users in the Amazon datasets have at least 5 interactions. To align with this setting, we filter the BeerAdvocate dataset using the same threshold. Due to unavailability of computational resources, we randomly sampled 25\% of users to form a BeerAdvocate subset.

Following the original setting of DeepCoNN~\citep{zheng2017joint} and its latest related research~\citep{DBLP:conf/sigir/SachdevaM20}, we use the Google News pre-trained word vectors~\citep{mikolov2013distributed} to generate pre-trained word embeddings. Each word in the review is thus represented as a 300-dimension vector.




We evaluate the rating prediction accuracy by computing the conventional Root-Mean-Square Error on the test set:
\begin{equation}
	rRMSE=\sqrt{\frac{\sum_{u,i}(r_{ui} -\hat{r}_{ui})^2}{N}} \label{RMSE}.
\end{equation}
where $N$ is the number of ratings. To evaluate recommendation performance for individual users, we also report per-user RMSE ($uRMSE$, as opposed to $rRMSE$) for further investigation. We cap the predicted ratings to $[1, 5]$, so there are no out-of-bounds values.


\subsection{Baselines}

We compare the performance of our proposed NAECF model with two related recommendation models:

\begin{itemize}
\item \textbf{Matrix Factorization~\citep{koren2009matrix}.} We use MF as a classical, pure similarity-based CF baseline. 
All non-textual hyper-parameters in NAECF are reused.
\item \textbf{DeepCoNN~\citep{zheng2017joint}.} This is the pioneering work and state-of-the-art method that introduces deep learning techniques to build a text-based recommender system.
User and item features are extracted in parallel, and their interaction is realized by means of factorization machines (FM). Although there are other text-based models following a similar architecture, such as NARRE~\citep{chen2018neural} and \citep{DBLP:conf/kdd/WangYWNHC19}, that may outperform DeepCoNN, the components they added for better recommendation performance are mainly attention layers or data augmentation modules, introducing no significant change in the model architecture. Therefore, to focus on the effect of autoencoders in NAECF, we still adopt DeepCoNN as the strongest and most relevant baseline. 
\end{itemize}
\subsection{Experimental Protocol}\label{sec:exp_pro}

We randomly split the datasets into training, validation and test sets with proportions 80\%, 10\% and 10\%, respectively. To address the influence of the data splitting strategy, we set 10 different random seeds and thus use 10 different splits. While all users have at least 5 interactions in total, a random split may distribute these interactions unevenly across sets, such that there may be users with only one rating in the training set. To address this potentially unreliable situation, we only account for users with at least 3 interactions in the training set for evaluation. 

We first do a grid search on the two Amazon datasets separately to fix the hyper-parameters on DeepCoNN. Then we reuse them for the investigation of NAECF. The hyper-parameters tuned are listed below, with the optimal values indicated in bold:
\begin{itemize}
    \item Number of latent factors for DeepCoNN and NAECF: $\{5, 10, 20, 50, \bm{100}, 200, 500\}$. All latent factors are initialized with a Uniform distribution between $-0.01$ and $0.01$.
    \item Learning rate: $\{\bm{0.00001}, 0.0001, 0.001, 0.01, 0.03, 0.1\}$.
    \item Dropout rate to avoid overfitting: $\{0, 0.1, \bm{0.2}, 0.5$\}.
    \item Batch size: $\{32, 64, 128, \bm{256}, 512, 1024\}$.
    \item Number of words: $\{128, 256, \bm{512}, 1024, 2048\}$.
    \item Length of CNN kernels: $\{2, \bm{3}, 4\}$.
\end{itemize}
 
Using the DeepCoNN architecture as reference, we investigate the impact of text reconstruction loss with different weights. Since our main concern in this paper is the effect of adding adversarial conditions via AEs to the original DeepCoNN setting, we weight the user and item AEs with the same weight $w$, as shown in Eq.~(\ref{eq:overall_loss}). Specifically, we consider weight values in the set $\{0, 0.1, 0.2, 0.5, 1, 2, 5, 10\}$. Note that NAECF reduces to DeepCoNN when $w=0$. Similar to the fine-tuning of the hyper-parameters, the optimal weight is selected on the validation set.

Autoencoders act as adversaries to the rating prediction process, so their activation may lower the overall validation $rRMSE$. Therefore, we deploy a two-stage training strategy: we set $w=0$ in the first 50 epochs as a pre-training process to get the model ready to train for NAECF, and then change $w$ to the value we are tuning for the next 50 epochs.

Since NAECF does not chase the best overall performance, but rather a better balance across users, we follow a different validation strategy for $w$. First, we separate users in bins based on their $uRMSE$ score with DeepCoNN; to stress more on the performance for non-mainstream users, we use the 4 uneven bins defined by percentiles 10, 50 and 90 of the $uRMSE$ distribution. The performance gain with respect to DeepCoNN is then computed using these bins as strata, assigning smaller importance to the first and last bins, that is, users with a good recommendation and users who are extremely difficult to model. This way, the assessment of model capability is better aligned with our purposes. Gain is thus defined as follows:
\begin{equation}
\label{eq:gain}
	\Delta=0.1\Delta_1+0.4\Delta_2 +0.4\Delta_3+0.1\Delta_4,
\end{equation}
where $\Delta_b$ indicates the mean $uRMSE$ difference between DeepCoNN and NAECF in user bin $b$, and bin weights reflect the fraction of users they contain out of the total sample. A positive $\Delta$ value means NAECF improves upon DeepCoNN. 
 
All models are implemented in PyTorch~\citep{paszke2017automatic}, with CUDA and CuDNN for acceleration on an NVIDIA GeForce GTX 1080Ti GPU. 


\section{Results}

\begin{table*}[t]
  \caption{$\bm{rRMSE}$ over 10 data splits for all recommendation models in all three datasets ($\bm{\mathit{mean} \pm \mathit{std.dev.}}$). Bold for best results per dataset. * for results statistically different from the best ($\bm{t}$-test, $\bm{p<0.05}$).}
  \label{tab:rmse}
  \centering{\footnotesize\setlength{\tabcolsep}{4pt}\begin{tabular}{|r|r|r|rrrrrrr|}
  \hline
  
  \multirow{2}{*}{Dataset} & \multirow{2}{*}{MF} & \multirow{2}{*}{DeepCoNN} & \multicolumn{7}{c|}{NAECF} \\ 
    & & & $w=0.1$ & $w=0.2$ & $w=0.5$ & $w=1.0$ & $w=2.0$ & $w=5.0$ & $w=10$.0\\
    \hline

  Instant Video  & 1.1600 $\pm$.0264* & 0.9744 $\pm$.0145 & \textbf{0.9732 $\pm$.0149} & 0.9749 $\pm$.0122 & 0.9754 $\pm$.0159 & 0.9757 $\pm$.0169 & 0.9798 $\pm$.0212 & 0.9896 $\pm$.0221* & 0.9967 $\pm$.0221*\\
  
  Digital Music & 1.0466 $\pm$.0097* & \textbf{0.9078 $\pm$.0138} & 0.9083 $\pm$.0115 & 0.9106 $\pm$.0128 & 0.9097 $\pm$.0114 & 0.9104 $\pm$.0128 & 0.9118 $\pm$.0134 & 0.9167 $\pm$.0108 & 0.9219 $\pm$.0146* \\
  
  BeerAdvocate & 1.0442 $\pm$.0048* & 0.6722 $\pm$.0090 & 0.6707 $\pm$.0064 & \textbf{0.6692 $\pm$.0035} & 0.6746 $\pm$.0059* & 0.6756 $\pm$.0082* & 0.6785 $\pm$.0098* & 0.6899 $\pm$.0137* & 0.7068 $\pm$.0278*\\  
  \hline
  \end{tabular}}
\end{table*}

In this section, we present and analyze the experimental results. As a summary, Table~\ref{tab:rmse} presents the mean performance of all models over the 10 splits per dataset. It can be seen that DeepCoNN and NAECFs show significantly better recommendation accuracy than MF (paired $t$-test, $p<0.05$~\citep{urbano2019statistical}), and that NAECF and DeepCoNN perform similarly overall, provided that the weight of the text reconstruction loss is not too high.
Furthermore, in Section~\ref{sec:results:rq2} we show that NAECF, while maintaining similar overall recommendation quality as DeepCoNN, manages to create a significantly better balance across users thanks to the introduction of the user and item reconstruction losses as adversaries to the rating prediction optimization.
Finally, in Section~\ref{seq:results:rq3} we dive deeper into the ability of the autocorrelates to reconstruct users from the learned representations, and how this correlates with the recommendation performance per user. This analysis sheds more light on the mechanics underlying NAECF and the reported results.



\begin{table}[t]
  \caption{Weights $w$ yielding the best performance gain per split on the validation set.}
  \label{tab:weights}
  \centering{\footnotesize\begin{tabular}{|r|rrrrrrrrrr|}
  \hline
  \multirow{2}{*}{Dataset} & \multicolumn{10}{c|}{Split} \\
     & 1 & 2 & 3 & 4 & 5 & 6 & 7 & 8 & 9 & 10 \\
    \hline
  Instant Video & 2 & 5 & 0.1 & 10 & 1 & 5 & 0.1 & 0.1 & 0 & 0.5 \\
  Digital Music & 0 & 0 & 5 & 5 & 2 & 0.1 & 2 & 0.1 & 0.1 & 0.5 \\
  BeerAdvocate  & 0.1 & 0.2 & 0 & 0.5 & 0 & 0.2 & 0.5 & 0.1 & 0.1 & 0.2 \\
  \hline
  \end{tabular}}
\end{table}

\subsection{Performance Balance Across Users} \label{sec:results:rq2}

In order to answer research questions RQ1 and RQ2, we investigate the effect of autoencoders and text reviews on the recommendations for non-mainstream and mainstream users. 

\subsubsection{Effect of autoencoders.} As an adversarial learning model, NAECF has two conflicting goals: minimizing the text reconstruction losses $L_U$ and $L_I$ versus minimizing the rating prediction loss $L_R$. If the weight of the text reconstruction loss is too small, autoencoders cannot exert sufficient influence on the training process, making them ineffective regarding the mainstream bias. Conversely, if the text reconstruction loss dominates the training process, we expect to have a significant drop in terms of overall rating prediction accuracy. 
Following the validation process in Section~\ref{sec:exp_pro}, we chose the weight $w$ with the best gain $\Delta$ on the validation set as the optimal one.


Table~\ref{tab:weights} reports the optimal validation-set weight per split.
As the table shows, in 5 of the 30 splits a weight $w=0$ achieved the best gain; note that such cases correspond to a simple re-training of DeepCoNN. However, in the vast majority of cases a weight different from zero yielded a better performance gain, although there does not appear to be a single optimal weight for the autoencoders in NAECF. Overall, this suggests that $w$ is a hyperparameter to tune on a case by case basis, and that autoencoders are expected to help when the characteristics of the data allow for it; sometimes they do not lead to a substantial gain over DeepCoNN. Furthermore, and based on detailed $rRMSE$ results not reported in the paper, we see that the weights with the best performance gains often lead to lower overall performance (7, 6, and 6 out of 10 seeds in three datasets). This contrast shows that a high score on an overall accuracy metric like $rRMSE$ does not necessarily reflect a good balance across individual users.

After the optimal weights are chosen on the validation set, we turn our attention to the corresponding test-set results. In Table ~\ref{tab:gains} we report the average performance gains of NAECF over DeepCoNN, both per bin and overall.
The table shows that users in the central bins (ie. central 80\% of users) receive a statistically significant performance gain on all datasets, which is exactly the users that we specifically target in NAECF.
For the two Amazon datasets, these are also the bins receiving the largest gains; for the BeerAdvocate dataset it is the first bin that has the highest gain, though the difference is not significant from the second bin. 
In fact, the gains and losses observed for the 10\% of users in the first bin are not statistically different from zero, which means that users that already receive good performance are neither helped nor punished by NAECF. Therefore, the application of autoencoders as adversaries to the rating prediction problem does not sacrifice performance for the mainstream users.
Finally, we observe that the 10\% of users in the last bin do receive a statistically significant performance loss. While unfortunate, such loss is a collateral damage on a minority of users who are hard to satisfy anyway, in benefit of the bulk of users who now receive better recommendations.
Averaging gains across bins, as indicated in Eq.~(\ref{eq:gain}), we see that NAECF yields statistically better results than DeepCoNN on all datasets. This indicates the overall success for NAECF to create a better balance across users. In general, we help most of the non-mainstream users without hurting mainstream users.
\begin{table}[t]
  \caption{Test-set performance gains averaged over splits (higher is better): per-bin gain $\bm{\Delta_b}$ and overall gain $\bm{\Delta}$. * for gains statistically different from 0 ($\bm{t}$-test, $\bm{p<0.05}$).}%
  \label{tab:gains}
  \centering{\footnotesize\begin{tabular}{|r|rrrr|r|}
  \hline
  Dataset & $\Delta_1$ & $\Delta_2$ & $\Delta_3$ & $\Delta_4$ & $\Delta$\\
  \hline
  Instant Video & -0.0035 & 0.0256* & {0.0267}* & -0.0308* & 0.0175*\\
  Digital Music & 0.0036 & {0.0184}* & 0.0106* & -0.0167* & 0.0103*\\
  BeerAdvocate  & {0.0119} & 0.0117* & 0.0063* & -0.0115* & 0.0073*\\
  \hline
  \end{tabular}}
\end{table}

\begin{figure}[t]
\centering\includegraphics[width=0.5\columnwidth]{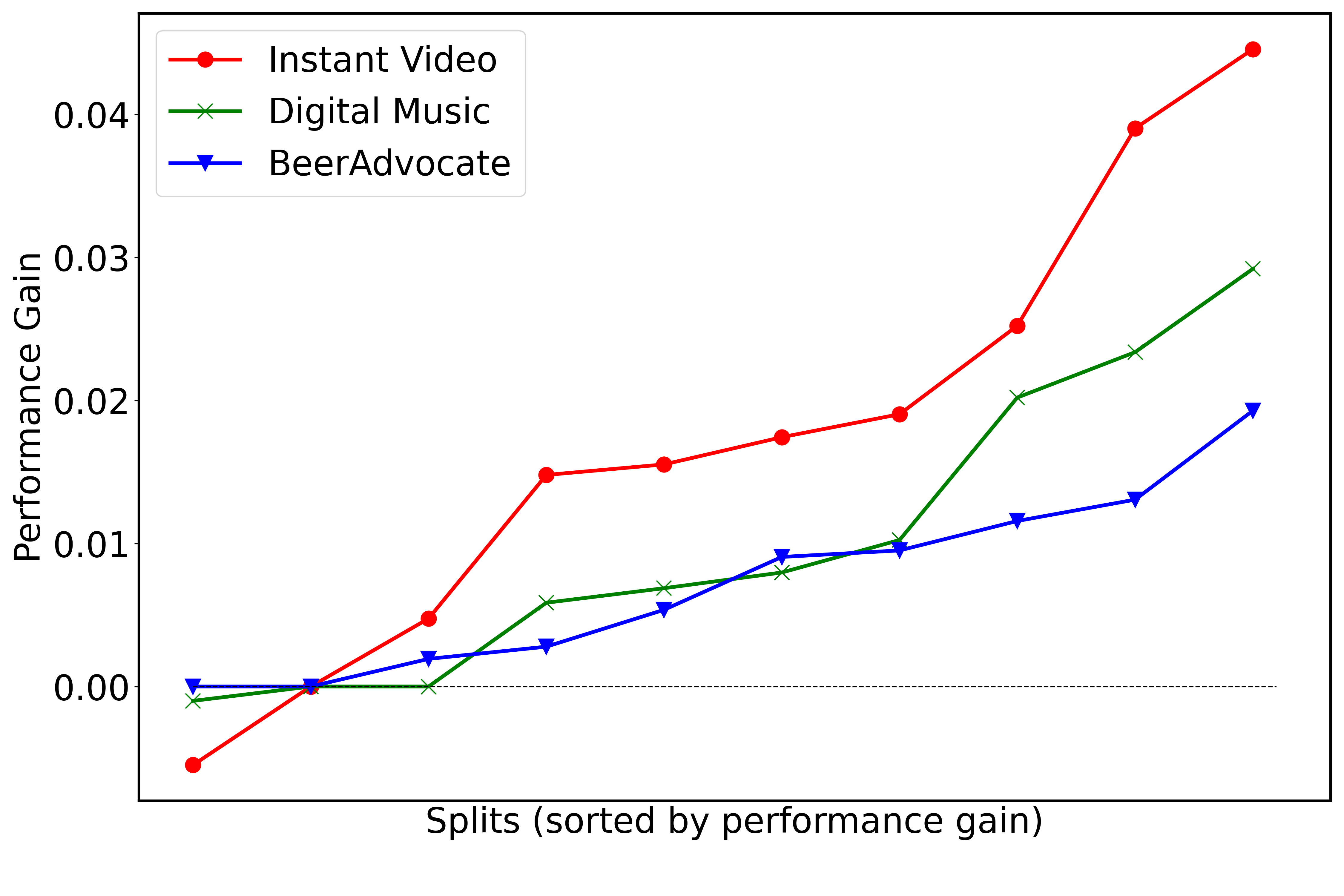}%
\caption{Test-set performance gain $\Delta$ on each of the 10 data splits, sorted within dataset.}%
\label{fig:q3}
\end{figure}

Figure~\ref{fig:q3} shows the test-set performance gain for the different data splits. We can first notice that the optimal weight, selected based on the gain on the validation set, turned into a slight loss in the test set for only one split in the Instant Video dataset ($\Delta=-0.0054$), and one split in the Digital Music dataset ($\Delta=-0.001$). Detailed results not reported in the paper show that this is mainly due to a drop in $\Delta_4$, representing the users that are hard to optimize for in any case.
In 5 cases the optimal weight was $w=0$, which yields a gain $\Delta=0$.\footnote{Due to the stochastic nature of the training process, one retraining of DeepCoNN may yield a slight gain with respect to another, but it should be zero on expectation. Therefore, we set $\Delta=0$ when $w=0$.}
For the majority of cases though (23 out of 30 splits), the optimal weight selection achieves a higher gain $\Delta$ and therefore helps achieve a better overall balance in recommendation performance across mainstream and non-mainstream users.
If we consider only top 90\% users to select the best weight, there are in total 28 out of 30 splits (except 2 in the Digital Music dataset) where NAECF shows superiority over DeepCoNN. However, this does not mean that a higher weight is always better for reaching a balance across users. For the BeerAdvocate dataset, 83\% of the top-3 weights selected via the validation process are not larger than 0.5. For the two Amazon datasets, and although the optimal weights distribute over all weight candidates, unreported results still show that weights no larger than 2.0 take 85\% of the top-3 best results. This observation matches our expectation that a mild weight value is more likely to bring a better trade-off between the overall recommendation accuracy and the balance across users. Based on these observations, we provide a positive answer to RQ1.

\subsubsection{Effect of text reviews.} We hypothesize that exploiting elaborate user- and item-related information, in our case in the form of online reviews, not only contributes to overall recommendation performance~\citep{zheng2017joint, chen2018neural}, but also to neutralizing the mainstream bias. While non-mainstream users are relatively underrepresented in the user space, it should at least help if their individual representations are as elaborate as possible to model their preferences better.

In order to verify this hypothesis, we investigate the effect of this additional information compared to the case where it is not used, such as in a classical collaborative-filtering models like MF.
We deliberately do not compare MF with NAECF because we would confound the use of text reviews for boosting recommendation accuracy and balancing across users. Instead, we choose to compare with DeepCoNN, which may anyway be regarded as a special case of NAECF and, architecture-wise it is the closest to collaborative filtering in the NAECF family. As such, a superiority of DeepCoNN over MF will indirectly mean a superiority of NAECF as well.

\begin{figure}[t]
\includegraphics[width=\columnwidth]{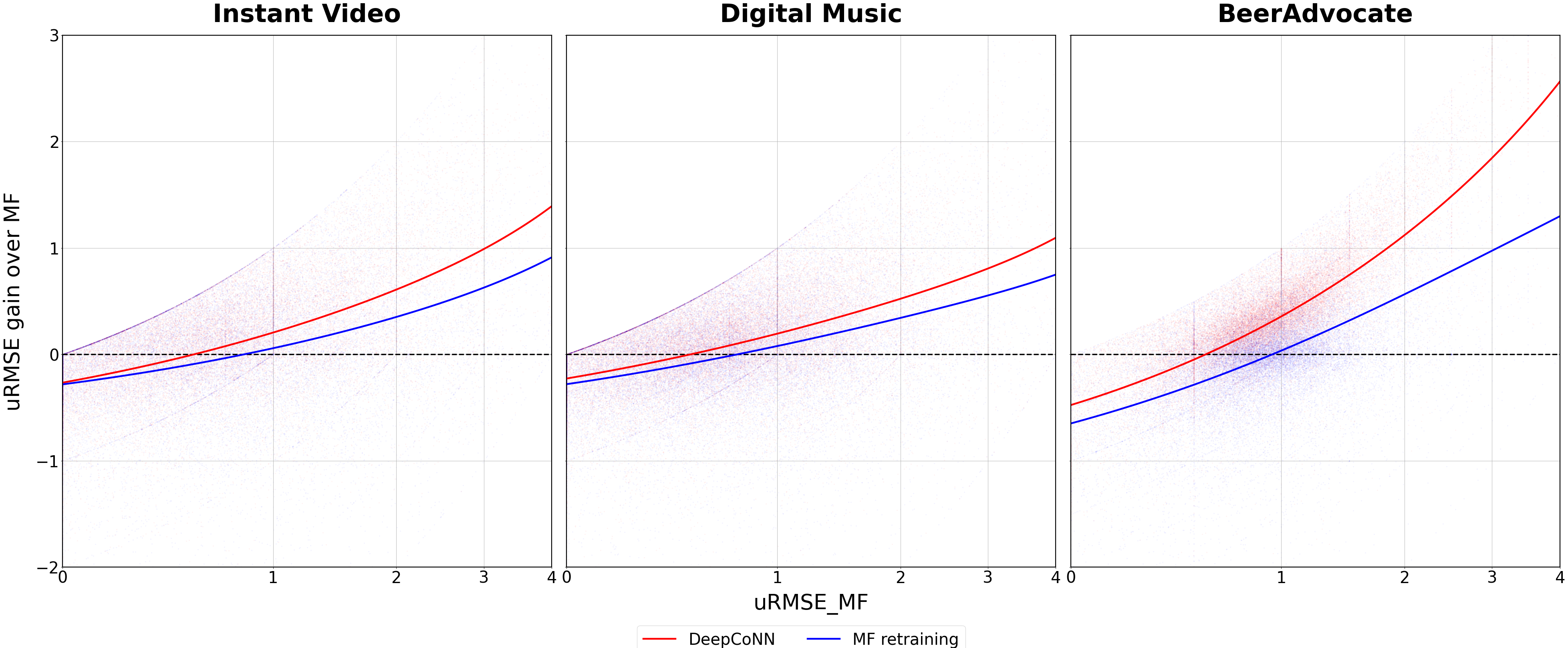}
\caption{$\bm{uRMSE}$ gain over MF (positive is better) of DeepCoNN and a retrained MF model, on all 10 data splits. Curves represent a spline-smoothed fit.}
\label{fig:q1}
\end{figure}
Fig.~\ref{fig:q1} shows the $uRMSE$ improvements on the test set made by DeepCoNN, compared to MF on all 3 datasets.
As the figure shows, our expectations are met on all three datasets. The improvement on $uRMSE$ scores has a clearly positive correlation with the baseline $uRMSE$ achieved by MF, meaning that it is the users who received worse recommendations in MF, the ones who benefit the most from the inclusion of textual features in DeepCoNN.
We note that, close to the origin of the plots, we see that DeepCoNN leads to slight performance loss for the users for which MF achieved the best performance. This is however an artifact of the evaluation process. Note that users with a $uRMSE$ close to zero in MF have almost no room for improvement, so any other model we compare with will probably perform worse. Similarly, other models will likely perform better for the users with very high $uRMSE$ in MF, because it is just not possible to perform worse.
To illustrate and account for this effect, Fig.~\ref{fig:q1} also compares with a retrained MF model, displaying both the overall correlation and the loss close to the origin. These serve as a sort of  baseline to assess the improvement of DeepCoNN (ie. rather than comparing the red curve with the $y=0$ axis, compare it with the blue curve). We can thus confirm the superiority of DeepCoNN, as $uRMSE$ scores are always higher than on a retrained MF model.

Finally, and similar to the comparison between DeepCoNN and NAECF stated in Eq.~(\ref{eq:gain}), here we also compare DeepCoNN and MF in terms of gain $\Delta$. 
The values on three datasets are 0.1175, 0.1017 and 0.3282, respectively. Such a significant improvement shows the effectiveness of review-based features in creating balance across different users, by which we provide an answer to RQ2. 

\vspace{1ex}\noindent In summary, we confirmed that NAECF creates a better balance across users by significantly improving the recommendation accuracy for non-mainstream users, subject to a good selection of the weight hyper-parameter. We also compared the review-based DeepCoNN and the CF-based MF, and found that the improvement stems mainly from a better optimization for non-mainstream users who are harder to handle in bare collaborative filtering. This way, NAECF's superiority lies in the use of review text, not only to boost rating prediction, but also as an adversary to ensure better user representation. Ultimately, these findings direct an open question to the correlation between the text reconstruction loss and the recommendation accuracy, which we study next.

\subsection{User Feature Reconstruction}\label{seq:results:rq3}


Mainstream users are generally active and display common behavioral patterns. This makes it easier for them to be matched with proper neighbors in collaborative filtering, ultimately giving them more accurate recommendations. At the same time, good performance on similarity-based user modeling will make them easier to reconstruct in the NAECF autoencoders, and should therefore have a lower text reconstruction loss after training.
This should be reflected by a positive correlation between $uRMSE$ scores and user reconstruction losses $loss_U$. Because DeepCoNN does not contain any text reconstruction module, the loss should be randomly distributed and uncorrelated with $uRMSE$; we verified this in the data but do not report it here. However, intuition tells us that mainstream users should have low reconstruction losses. The failure of DeepCoNN to reflect this expectation means there is room for improvement to create a better balance across users, which is confirmed by our findings in the previous section.

\begin{figure}[t]
\centering\includegraphics[width=.91\columnwidth]{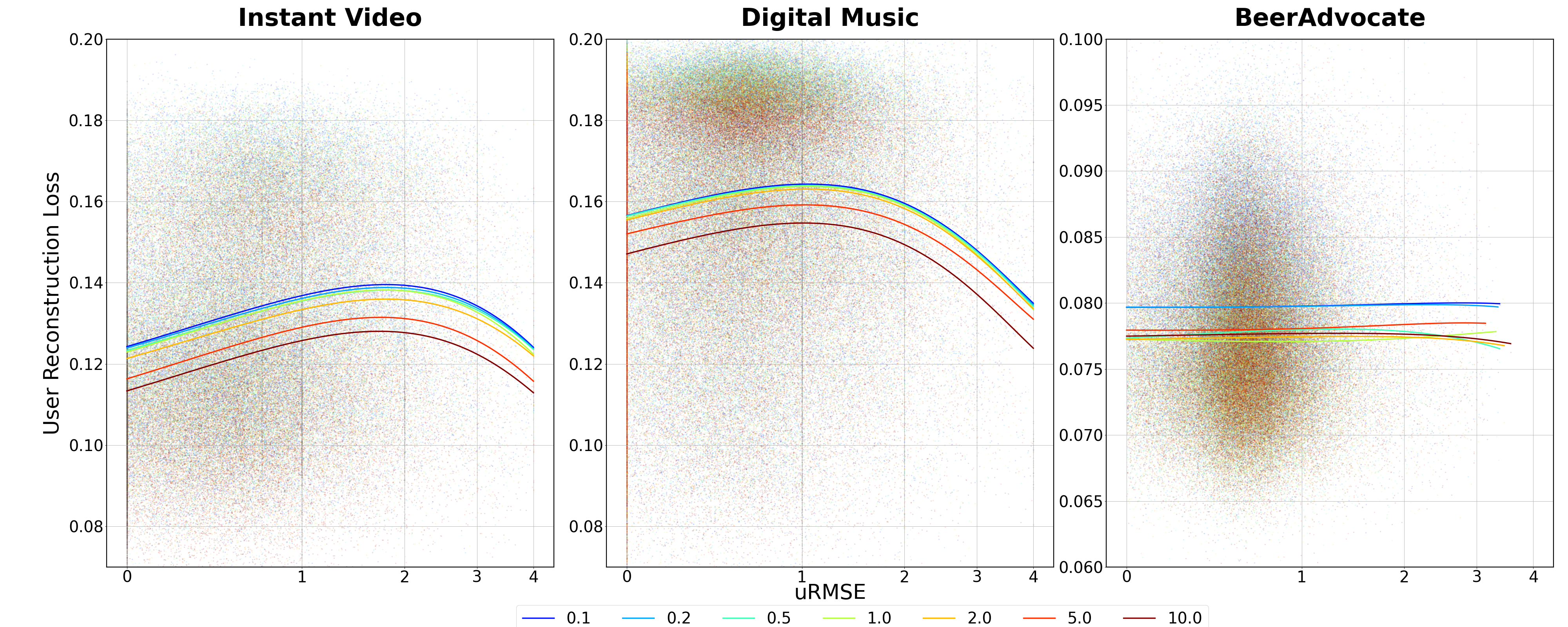}
\caption{NAECF $\bm{loss_U}$ by test-set $\bm{uRMSE}$, for each weight $w$. Lines represent a spline-smoothed fit.}
\label{fig:rmse_cos}
\end{figure}

Therefore, we now look into the correlation between user reconstruction loss $loss_U$ and $uRMSE$ recommendation accuracy. Fig.~\ref{fig:rmse_cos} shows the relationship for each of the evaluated weights $w$. We can see clear differences across datasets, but there are several qualitative commonalities. First, the majority of users are not mainstream even if they have a rather low $uRMSE$ score. Second, higher weights generally lead to lower reconstruction losses and therefore to better user representations. This is expected because a high weight makes the text reconstruction losses dominate the overall loss in Eq.~(\ref{eq:overall_loss}), but the figure further shows that the relative relationship between $loss_U$ and $uRMSE$ is pretty consistent across weights. Interestingly, the BeerAdvocate dataset shows some fluctuations with high weights. This evidences that the optimal weight needs proper tuning, because excessively high weights lead to substantial performance loss and $uRMSE$ scores become less stable as a consequence. As reported in Table~\ref{tab:weights}, the optimal weights for this dataset are rather small indeed in comparison with the other two datasets.

\begin{figure}[t]
\centering\includegraphics[width=\columnwidth]{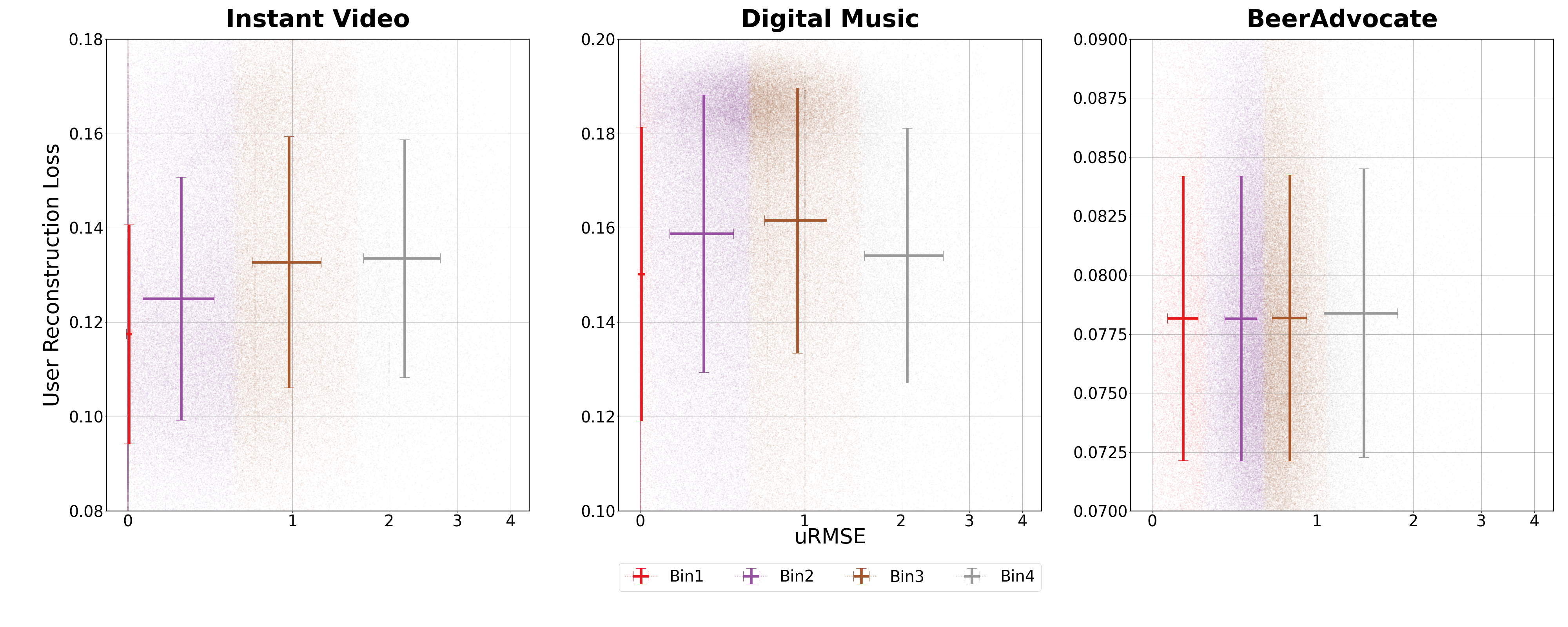}%
\caption{NAECF $\bm{loss_U}$ by test-set $\bm{uRMSE}$, with all evaluated weights $w$. Error bars show standard deviations per user bin.}
\label{fig:bin}
\end{figure}

We also followed the earlier approach of dividing users in four bins according to the $uRMSE$ distribution. Fig.~\ref{fig:bin} similarly shows the relationship between $loss_U$ and $uRMSE$ with  all the evaluated weights, but differentiating among user bins. We can clearly observe that, as expected, the relationship is monotonically positive except for the last bin in the Digital Music dataset. This confirms again that users who are better represented receive more accurate recommendations.
As reported in Table~\ref{tab:gains}, mainstream users in bin 1 do not always benefit from NAECF because they already receive good recommendations and there is little room for improvement, regardless of how well they are reconstructed. Fig.~\ref{fig:bin} confirms this specially in the two Amazon datasets, where bin 1 users receive nearly perfect recommendations. But NAECF improves performance specially for the 80\% of non-mainstream users in bins 2 and 3, because those are harder to represent to begin with. Fig.~\ref{fig:bin} confirms that these users generally have the highest reconstruction losses indeed. Together with the correlations in Fig.~\ref{fig:rmse_cos}, we see the relationship between the mainstreamness of users and the difficulty to represent them.
Notwithstanding, the bottom 10\% of users in bin 4 are too extreme to find proper representations, so the autoencoders hardly work for them. We even observed in Fig.~\ref{fig:bin} a negative correlation on the Instant Video and Digital Music datasets when $uRMSE$ is large. This confirms that NAECF sacrifices performance for these extreme users in favor of the others. Although unfortunate, we find this behavior acceptable because these users often display such particular tastes and patterns that it is hard for them to benefit from virtually any CF-based recommendation model.

\section{Conclusion and Future Work}

Rating accuracy has long been an important criterion to evaluate recommender systems, if not the most important. Previous research has therefore focused mainly on maximizing the overall performance averaged over users. However, traditional collaborative filtering methods focus more strongly on recommending items that have positive interactions by similar users. In this situation, it is hard for CF models to work well with non-mainstream users that have special tastes or habits.
Because non-mainstream users are rather a minority, this problem may not have a strong effect on the overall accuracy, yet it may create an unfair imbalance across users. To address this problem, we proposed a conceptually simple but effective model named NAECF, which minimizes the rating prediction loss while keeping the user and item properties preserved in the learned user and item representations. Preservation of user and item properties is imposed as an adversarial condition by minimizing reconstruction losses in addition to rating prediction error. This prevents these representations from being biased towards mainstream users.

We conducted experiments on three real-world datasets, and found that NAECF achieves an overall rating accuracy that is on par with the state-of-the-art. However, its strength is in the better balance it achieves across users thanks to a significant improvement of the recommendation accuracy for non-mainstream users, without significantly harming the mainstream ones. This improvement is achieved through an optimal trade-off between rating prediction and text reconstruction. Our results confirm a clear correlation between how well users are represented and the quality of their recommendations, evidencing that side information may be instrumental not only for boosting overall accuracy, but also to minimize possible biases in the learned models.



Future work will be conducted in several directions. First, we will investigate whether the conclusions drawn here for rating prediction generalize to the ranking paradigm, which is gaining popularity in the recommendation field. Second, in this paper we treated users and items as equally important through a single text reconstruction weight. One may argue that improving the representation of users alone is not enough, because the model also needs a good item representation to know what to recommend. However, users and items may have different impacts, and we would like to explore this question by implementing two weights in the NAECF loss. Third, we introduced side information from text reviews in order to achieve a better balance across users. However, text reviews are just an example of additional content-based resources such as images and demographic information that can be used to achieve a similar function. We would like to further investigate the effect of other side information in the future and, perhaps more importantly, how to effectively incorporate such information in NAECF to eliminate the mainstream-bias. Finally, we are also interested in combining NAECF with explainable recommendation, so that we can provide convincing explanations to non-mainstream users.


%
\bibliographystyle{ACM-Reference-Format}
\bibliography{sample-base}

%

\end{document}